# The Amplitude Modulation Structure of Japanese Infant- and Child-Directed Speech: Longitudinal Data Reveal Universal Acoustic Physical Structures Underpinning Moraic Timing

Tatsuya Daikoku [1,2] and Usha Goswami [1]

[1] Centre for Neuroscience in Education, University of Cambridge

[2] Graduate School of Information Science and Technology, The University of Tokyo, Tokyo, Japan

**Author Note**

Correspondence concerning this article should be addressed to Tatsuya Daikoku, Graduate School of Information Science and Technology, and The University of Tokyo, Tokyo, Japan Centre for Neuroscience in Education, University of Cambridge, Cambridge CB2 3EB, United Kingdom. Email: td441@cam.ac.uk daikoku.tatsuya@mail.u-tokyo.ac.jp

## Competing Interests

The authors declare no competing financial interests.

## Author Contributions

T.D. and U.G. conceived the method of data analysis. T.D. organized the corpus data. T.D. analysed the speech waveform and conducted statistical analyses, and summarized all the results. T.D. wrote the draft of the manuscript and figure. U.G. and T.D. edited and finalized the manuscript.

## Acknowledgements

This research was supported by JSPS KAKENHI (Grant Number 24H00898) and The National Institute for Japanese Language and Linguistics (NINJAL), Japan, to T.D. and by a donation from the Yidan Prize Foundation to U.G. The funders had no role in study design, data collection and analysis, decision to publish, or preparation of the manuscript.

## Data Availability

In this paper, we used "NTT Infant Speech Database (INFANT)" provided by Speech Resources Consortium, National Institute of Informatics (https://doi.org/10.32130/src.INFANT). All of anonymized data files analyzed have been deposited to an external source (https://osf.io/n3upf/). The other data and results of statistical analysis are shown in supplementary data.

# Abstract

Infant-directed speech (IDS) is highly rhythmic, and in European languages IDS is dominated by patterns of amplitude modulation (AM) at ~2Hz (reflecting prosody) and ~5Hz (reflecting individual syllables). The rhythm structure of spoken Japanese is thought to differ from European stress-timed and syllable-timed languages, depending on moraic units. Morae comprise any onset phoneme and vowel phonemes within a syllable, PA-N-DA. Arguably, initial speech encoding via cortical tracking by infants is likely to utilize language-universal physical acoustic structures in speech rather than language-specific structures like morae, since the infant brain must be prepared to acquire any human language. Here a language-blind computational model of linguistic rhythm based on features of the amplitude envelope is used to compute these physical acoustic stimulus characteristics for Japanese. Using ~18,000 samples of natural IDS and child-directed speech (CDS) recorded longitudinally from 6 parents while speaking to their children over the ages 0 – 5 years, we find that the temporal modulation patterns that characterise the amplitude envelope of Japanese are highly similar to those found for stress-timed and syllable-timed European languages. However, the AM band corresponding to the syllabic level in CDS/IDS in European languages (~2-12Hz, theta-rate cortical tracking) was elongated in Japanese (2.5-17Hz). Further, the phase synchronization ratios between the two slowest AM bands were as likely to be 1:3 as 1:2, which differs from European languages where 1:2 ratios are dominant. Accordingly, the language-universal amplitude-driven physical acoustic structures important for cortical speech tracking flexibly accommodate language-specific differences in core rhythmic units.

*Keywords:* Rhythm, Speech, Phonological hierarchy, Amplitude modulation

# 1. Introduction

How infants, born without understanding or producing words, acquire spoken language is a long-standing research puzzle. A central piece of this linguistic puzzle is infant-directed speech (IDS)—the distinctive, melodic form of speech that caregivers instinctively use when speaking to babies. From their first moments of birth, infants are surrounded by IDS, and the rich auditory and rhythmic cues provided by IDS are thought to guide their earliest steps into language development. Caregivers unconsciously use IDS when interacting with their infants, and IDS has a number of defining characteristics including a simplified grammatical structure, an exaggerated rhythmic structure, heightened pitch, exaggerated pitch range and hyperarticulation of vowels, a feature that serves to exaggerate phonetic categories (Fernald et al., 1989; Fernald & Simon, 1984; Kuhl et al.,1997, 2000, 2004; Soderstrom, 2007; Leong et al., 2017). Yet despite decades of research, much remains unknown concerning the neural and perceptual mechanisms that infants use to profit from the special characteristics of IDS to learn language.

In the current report, we adopt an oscillatory framework derived from recent research on the neural mechanisms of adult speech encoding and investigate whether early language acquisition may be grounded in the internalization of the rhythmic structures that exist across different languages. We use this neural perspective to motivate speech modelling of the amplitude envelope (AE) of Japanese IDS and child-directed speech (CDS). The AE is the slow-varying energy profile of the acoustic waveform that is known to be tracked by cortical oscillations (Giraud & Poeppel, 2012), and captures variations in signal intensity (amplitude modulation, AM) over time (Houtgast & Steeneken, 1985). AE-based speech modelling has shown that French and German poetry, which are classically considered to utilize fundamentally different linguistic structures to create rhythmic regularity, are actually remarkably similar at the level of their acoustic AM structures (Daikoku, Lee & Goswami, 2024). AE-based modelling has also shown high similarity regarding the acoustic physical structures of CDS in stress-timed (English) versus syllable-timed (Spanish) languages (Leong & Goswami, 2015; Perez-Navarro et al., 2022). Both English and Spanish CDS are dominated by AMs centred on ~2Hz and ~5Hz, with the phase alignment between these two AM bandings oscillating in a predominantly 1:2 ratio and yielding the perception of rhythm (Leong et al., 2014). These AM bandings are known to be tracked cortically by neuroelectric oscillations operating at matching rhythmic rates, delta-band networks (0.5-4 Hz) and theta-band networks (4-8 Hz, Poeppel, 2014). Given that neonates are sensitive to speech rhythm (Mehler et al., 1988; Nazzi et al., 1998), these acoustic AM-driven rhythmic structures may serve as a foundational perceptual scaffold for language development. Accordingly, here we apply the same AE-based modelling previously applied to English and Spanish IDS and CDS to IDS and CDS in Japanese. If the oscillatory framework of neural speech encoding is language-universal, as could be expected given that the biophysical and neurophysiological constraints upon IDS speakers are highly similar, then Japanese IDS and CDS could be expected to have the same rhythmic "beat" structures of European stress-timed (English) and syllable-timed (Spanish) languages, with dominant “delta-rate” and “theta-rate” AM bandings characterized by a 1:2 phase synchronization ratio.

However, speech rhythm in Japanese is thought to be fundamentally different from the rhythm structures that characterize European languages, with a dominant ~10Hz rhythm (Peter et al., 2022). Japanese spoken language is recognized for employing a mora-timed rhythmic structure, which sets it apart from the stress-timed rhythms of languages like English and German and the syllable-timed rhythms of languages like French and Spanish. Mora-timed languages, such as Japanese, Gilbertese, Slovak, and Ganda, are unique in using morae—sub-syllabic units with uniform temporal duration—as the foundational timing unit. For example, a word like PANDA would be segmented into the morae PA-N-DA (Peter et al., 2022).This rhythmic typology is thought to reflect a temporal organization of units occurring at frequencies around 8-10Hz (Han, 1994; Warner & Arai, 2001). However, these acoustic analyses are based on adult-directed speech (ADS). It is possible that regarding IDS and CDS, which have a language-teaching function, more similar rhythmic beat structures will be found as identified in stress-timed and syllable-timed European languages, with phase relations between delta-rate and theta-rate AM bandings yielding the perception of speech rhythm. Given that the infant brain needs to be prepared to learn any human language, whatever the characteristics of its rhythmic timing, IDS in all languages may be rhythmically extremely similar at the level of acoustic physical structures.

Accordingly, the current study analyzed the acoustic structure of Japanese ADS, CDS and IDS from a language-blind AE-perspective. By utilizing a longitudinal corpus previously used to demonstrate exaggerated pitch (Amano, Nakatani & Kondo, 2006) and vowel hyperarticulation in Japanese IDS (Ishizuka et al., 2007), here we provide complementary analyses of the acoustic structure of the AE of Japanese IDS/CDS relevant to cortical speech encoding (Giraud & Poeppel, 2012). Our acoustic modeling approach focuses on the AM hierarchy of sound waveforms below approximately 40 Hz, which characterise the speech signal as processed by human listeners (Daikoku & Goswami, 2022). These different rates of AM arise from the coordinated movements of the vocal folds, tongue, vocal tract, mouth and lips (Greenberg, 2003), which collectively generate a varying AE underpinning the perception of speech rhythm (Greenberg, 2006). Indeed, these motor movements exhibit a high degree of temporal consistency across languages (Poeppel & Assaneo, 2020). Each AM band is associated with specific linguistic units. Delta-rate modulations (<4 Hz) facilitate prosodic parsing, theta-rate modulations (4–12 Hz) support syllabic parsing, and beta/low gamma-rate modulations (12–40 Hz) aid in identifying phoneme onsets within syllables (Leong & Goswami, 2015). The terminology for these bands—delta, theta, beta, and gamma—derives from neurophysiology, as neural cell networks oscillating at these frequencies encode this AM information (Gross et al., 2013).

Our primary goal was to determine whether the acoustic physical structure of Japanese speech reveals nested hierarchical temporal modulation patterns in the AE that align with other languages. To achieve this, we employed the Spectral-Amplitude Modulation Phase Hierarchy (S-AMPH) model (Leong, 2012; Leong & Goswami, 2015), a computational framework designed to analyze the AM structure of the AE by isolating AM characteristics from frequency modulation (FM) components. The S-AMPH model provides a low-dimensional representation of the speech signal, capturing its dominant spectral (acoustic frequency, including pitch and formants) and temporal (oscillatory rate and speech rhythm) modulation patterns. This method enables the identification of predominant timescales of rhythmic modulation within Japanese speech. Furthermore, prior S-AMPH studies have demonstrated that IDS and music provide more rhythmically structured input than natural ADS, as evidenced by stronger phase relationships between the delta and theta AM bands and a prominent 1:2 ratio as regards the phase synchronization index (Daikoku & Goswami, 2022; Leong et al., 2017; Perez-Navarro et al., 2022). The phase relations between different AM bandings in Japanese IDS/CDS and Japanese ADS are hence also investigated.

Building on previous S-AMPH modelling in other languages, we hypothesize that the AM bands and their boundaries—defining the temporal modulation structure of spoken Japanese—will closely align with the AM bands identified in other languages. These prior languages are English (IDS, CDS and ADS), French (adult poetry), German (adult poetry), Portuguese (ADS), and CDS in Spanish (Leong & Goswami, 2015; Leong et al., 2017; Araujo et al., 2018; Daikoku & Goswami, 2022; Perez-Navarro et al., 2022; Daikoku et al., 2024). We also predict a comparable AM phase relationship pattern for Japanese IDS/CDS as found in these languages, with the delta- and theta-rate AM bands showing a higher phase synchronization index (PSI) in IDS/CDS than in ADS. A PSI of 1:2 would signal the presence of a distinct rhythmic beat, reinforcing the universality of rhythmic temporal structures in spoken language.

The longitudinal database employed here comprised naturalistic speech recordings from three families, 6 parents and 5 children. The speech interactions were recorded intermittently over a period of up to five years from the time of each infant's birth (https://doi.org/10.32130/src.INFANT; Amano et al., 2008). The consequent database was much larger than that used in the first CDS studies in English (6 speakers, who each recited 44 nursery rhymes yielding 264 speech samples for analysis, Leong & Goswami, 2015). Leong and Goswami's (2015) modelling identified a temporal modulation structure since replicated for English IDS (24 speakers, Leong et al., 2017) and Spanish CDS (18 speakers, Perez-Navarro et al., 2022). Longitudinal databases based on extended recording periods are exceptionally rare. In contrast, cross-sectional databases—constructed by aggregating short-term recordings from multiple infants of varying ages—are relatively common. This discrepancy arises primarily from the logistical challenges inherent in longitudinal data collection. Recruiting participants for short-term recordings, as required for cross-sectional databases, is considerably easier compared to securing long-term commitments necessary for longitudinal studies. Longitudinal data enable detailed observations of the emergence and acquisition sequence of specific linguistic abilities. These insights can be instrumental in testing existing theories of language development and constructing robust developmental models.

# 2. Materials and Methods

## 2.1. Participants

This study utilized a longitudinal database of spontaneous speech recorded within the home environment, encompassing three families with a total of five young children and their parents. These recordings were conducted intermittently over a period of up to five years, beginning shortly after the children's birth (https://doi.org/10.32130/src.INFANT) (Amano et al., 2008). All five children were born and raised in urban areas where standard Japanese, with minimal influence from regional dialects, is predominantly spoken. Both the children and their parents were healthy, with no reported abnormalities in speech perception or production.

## 2.2. Procedure

### 2.2.1. Recordings

The database used in this study consists of recordings captured using a DAT recorder (TCD-D10, SONY) and its stereo microphone (ECM-959, SONY) in mono format, with a 16-bit resolution and a sampling rate of 16 kHz. Recording sessions were conducted approximately once a month, whenever the children were in a good mood, to ensure high-quality, naturalistic speech samples. This resulted in >9,700 recordings for analysis regarding maternal IDS/CDS, and >8,300 recordings for analysis regarding paternal IDS/CDS. The microphone was either handheld by a parent or mounted on a microphone stand. While the majority of recordings took place within the children's homes, occasional sessions occurred in alternative settings, such as hospitals, the parents' hometowns, or vacation accommodations. To prioritize capturing naturalistic speech interactions, no specific tasks or structured activities were imposed during the recordings.

Recordings commenced within one month of the children's birth. However, due to familial circumstances, recordings were skipped in certain months, and the duration of recordings varied between months. Detailed information on the recording dates and durations is shown in Table 1 and available through an external source (https://osf.io/n3upf/).

### 2.2.2. Preprocessing of the raw data

Speech data were extracted from the recordings that ranged from approximately 15 minutes to 1 hour per session. During this extraction process, speech data were segmented based on their context. Temporally adjacent utterances from the same speaker with silent intervals shorter than 500 ms were merged into a single speech file. Conversely, when silent intervals exceeded 500 ms, the speech data were treated as separate events, resulting in two distinct files. This procedure ensured the creation of multiple speech files for each session, accurately reflecting the natural flow of conversation.

To ensure reliable results of frequency analysis, particularly for examining the minimum target frequency of 0.9 Hz in the S-AMPH model used in this study, each speech file required sufficient length. Files shorter than 10 seconds were merged with temporally adjacent files from the same session to create a file exceeding 10 seconds in length. To maintain the integrity of individual utterances, 1 second of silence was inserted between the concatenated files. Consequently, all merged data files ranged in length from 10.1 to 12.5 seconds. Table 1 provides the mean length of the processed speech files. Then, for each speech file longer than 10 seconds, 3 seconds of silence were added to both the beginning and end of the speech file to ensure robust analysis across the entire speech waveform.

To verify that file lengths did not significantly differ between groups—child-directed speech by mother (CDSm) and father (CDSf), adult-directed speech by mother (ADSm) and father (CDSf)—we first performed the Shapiro–Wilk test for normality on the length values in each group. Depending on the result of the test for normality, either the parametric or non-parametric (Kruskal-Wallis) analysis of variance (ANOVA) including a between-group factor (CDSm, CDSf, ADSm, ADSf) and a between-group factor (subject A, B, C, D, and E) was applied. Statistical analyses were conducted using jamovi Version 1.2 (The jamovi project, 2021). We selected $p < .05$ as the threshold for statistical significance. The results indicated significant differences in file lengths among the groups ($p < .05$, in each group) (https://osf.io/n3upf/). However, all effect sizes were exceptionally small (ranging from 0.00389 to 0.02974), suggesting that the observed significance is likely attributable to the large sample size rather than substantive differences between groups. Actually, all merged data files ranged in length from 10.0 to 12.5 seconds in each group (Table 1). This means no large difference in the length of speech data. A detailed summary of the speech file lengths, including mean values and standard deviations, is presented in Table 1. Following preprocessing, the speech data were analyzed to investigate the amplitude modulation (AM) structures in the amplitude envelope (AE) of the speech waveforms.

**Table 1. Information of the Speech Sample**

| ID | Sex | At birth | | | Recording periods | Mean length (sec±SD) | | | |
|---|---|---|---|---|---|---|---|---|---|
| | | Birth day | Height (mm) | Weight (g) | | ADSm | ADSf | CDSm | CDSf |
| A | M | 1988/4 | 510 | 3450 | 1988/5~1990/10 | 10.7(±0.14) | 10.9(±0.22) | 10.3(±0.03) | 10.4(±0.15) |
| B | M | 1990/5 | 480 | 3250 | 1990/6~1994/11 | 11.2(±0.19) | 11.1(±0.17) | 10.5(±0.03) | 10.4(±0.03) |
| C | F | 1991/6 | 495 | 3440 | 1991/6~1996/6 | 11.8(±0.34) | 10.8(±0.29) | 10.2(±0.01) | 10.1(±0.02) |
| D | F | 1994/8 | 485 | 3224 | 1994/8~1999/8 | 11.4(±1.11) | 12.0(±1.22) | 10.2(±0.03) | 10.0(±0.01) |
| E | F | 1995/2 | 502 | 3245 | 1995/2~2000/2 | 12.5(±0.42) | 11.5(±0.30) | 10.5(±0.05) | 10.6(±0.11) |

* M = male, F = female, CDSm = child-directed speech by mother, CDSf = child-directed speech by father, ADSm = adult-directed speech by mother, ADSf = adult-directed speech by father

### 2.3. Data Analysis

The comprehensive modeling framework developed by Leong and Goswami (2015; for further details, see https://www.cne.psychol.cam.ac.uk) offers a robust methodology for capturing the hierarchical amplitude modulation (AM) structures across diverse speech styles. This innovative approach involves generating high-dimensional representations of the acoustic signal, which are subsequently subjected to principal component analysis (PCA) to disentangle the spectral and temporal dimensions of the signal (referred to as spectral PCA and temporal PCA, respectively). Through this process, the model identifies the fundamental modulation bands characteristic of each speech register, providing the underlying acoustic architecture (Figure 1).

#### *2.3.1. Spectral Amplitude Modulation Phase Hierarchy (S-AMPH)*

To exclude the potential effects of sound intensity on spectrotemporal modulation features, the acoustic speech signals were z-score normalized (M = 0, SD = 1). The raw acoustic signals were processed through a 28-channel logarithmically spaced $ERB_N$ filter bank, spanning a frequency range of 100–7250 Hz. This filter bank emulates the cochlear frequency decomposition observed in a typical human auditory system, enabling the identification of spectral modulation patterns (Dau et al., 1997; Moore, 2012). Technical specifications of the filter bank design are detailed in Stone and Moore (2003), with additional parameter settings and frequency response characteristics provided in Appendix S1. For each of the 28 frequency channels, signal envelopes were derived using the Hilbert transform, resulting in a set of 28 Hilbert envelopes. The PCA was then applied to these envelopes, uncovering the core spectral modulation patterns by identifying the number and spacing of non-redundant spectral bands based on co-modulation within this high-dimensional $ERB_N$ representation. Temporal modulation patterns were subsequently extracted by filtering the raw acoustic signal into the spectral bands identified through the spectral PCA analysis. This two-step process establishes a precise and comprehensive framework for characterizing both spectral and temporal modulation features.

In the spectral PCA, analysis was focused on the top five principal components (PCs) (Figure 1, middle), which collectively explained over 70% (PC1: 46.16%, PC2: 10.45%, PC3: 6.32%, PC4: 4.78%, and PC5: 3.86%) of the total variance in the original acoustic signal on average. To explore temporal modulation patterns, the raw acoustic signal was filtered into five distinct spectral bands derived from the spectral PCA. Subsequently, the AM envelope within each spectral band was processed through a 24-channel logarithmically spaced $ERB_N$ filter bank covering the frequency range of 0.9–40 Hz (see Appendix S2 for further details). The Hilbert transform was applied to each of the 24 filtered signals to extract their envelopes. Temporal PCA was then performed to uncover the core temporal modulation structures. Additionally, mean frequency power (MFP) was analyzed to identify which spectral bands made the most substantial contributions to the temporal amplitude structure. For the temporal PCA, only the top three PCs were retained for further analysis, as they cumulatively accounted for over 90% (PC1: 71.38%, PC2: 13.00%, and PC3: 4.68%) of the variance in the original signal (Figure 1, right). Detailed PCA results can be found in Appendix S3. To identify the hierarchical modulation patterns, the absolute values of PC loadings were averaged across all speech samples. Peaks in the grand average PC loading patterns were identified to determine the dominant modulation clusters, while troughs were examined as they delineate the boundaries between co-modulated channel clusters (Daikoku & Goswami, 2022). This comprehensive approach highlights the hierarchical organization of spectrotemporal modulation in speech signals.

To identify the number and boundaries of the core modulation bands in both the spectral (acoustic frequencies spanning 100–7,250 Hz) and temporal (oscillatory rates spanning 0.9–40 Hz) domains, the PCA was applied to reduce dimensionality in each domain. PCA, a well-established method for dimensionality reduction in speech research (e.g., Klein et al., 1970; Pols et al., 1973), was employed to uncover the underlying structure of modulation patterns. This analysis focused on the absolute values of component loadings rather than component scores. Component loadings provide insight into the patterns of correlation between high-dimensional channels, reflecting how these channels co-vary. Specifically, PCA loadings were used to identify clusters of co-modulated channels within the spectral (28 channels) and temporal (24 channels) domains, thereby delineating the core modulation bands. These clusters represent the fundamental modulation structures underlying the speech signal. Detailed methodology and additional information on the PCA approach are provided in Appendix S1.

To identify the core spectral and temporal modulation bands, we established criteria grounded in prior research utilizing the S-AMPH framework (Leong et al., 2015; Daikoku & Goswami, 2022). Specifically, to ensure sufficient separation between the inferred modulation bands, a minimum peak-to-peak distance of two channels for spectral PCA and five channels for temporal PCA was imposed. This spacing criterion aimed to prevent over-fragmentation of modulation patterns and to reflect meaningful structural organization in speech signals. Additionally, cycles in which the peak-to-peak amplitude was less than 10% of the average peak-to-peak amplitude were excluded, as these were deemed to represent noise rather than meaningful speech cycles. Peaks and troughs were systematically identified across all PCs, with troughs representing the boundaries between modulation bands. These peaks signify the presence of co-modulated clusters, which constitute the primary modulation structures. Boundary edges between bands were delineated based on the most consistent locations of "flanking" troughs, which demarcated the limits of each modulation band. This rigorous approach ensured a robust and interpretable characterization of the spectral and temporal modulation hierarchy in speech.

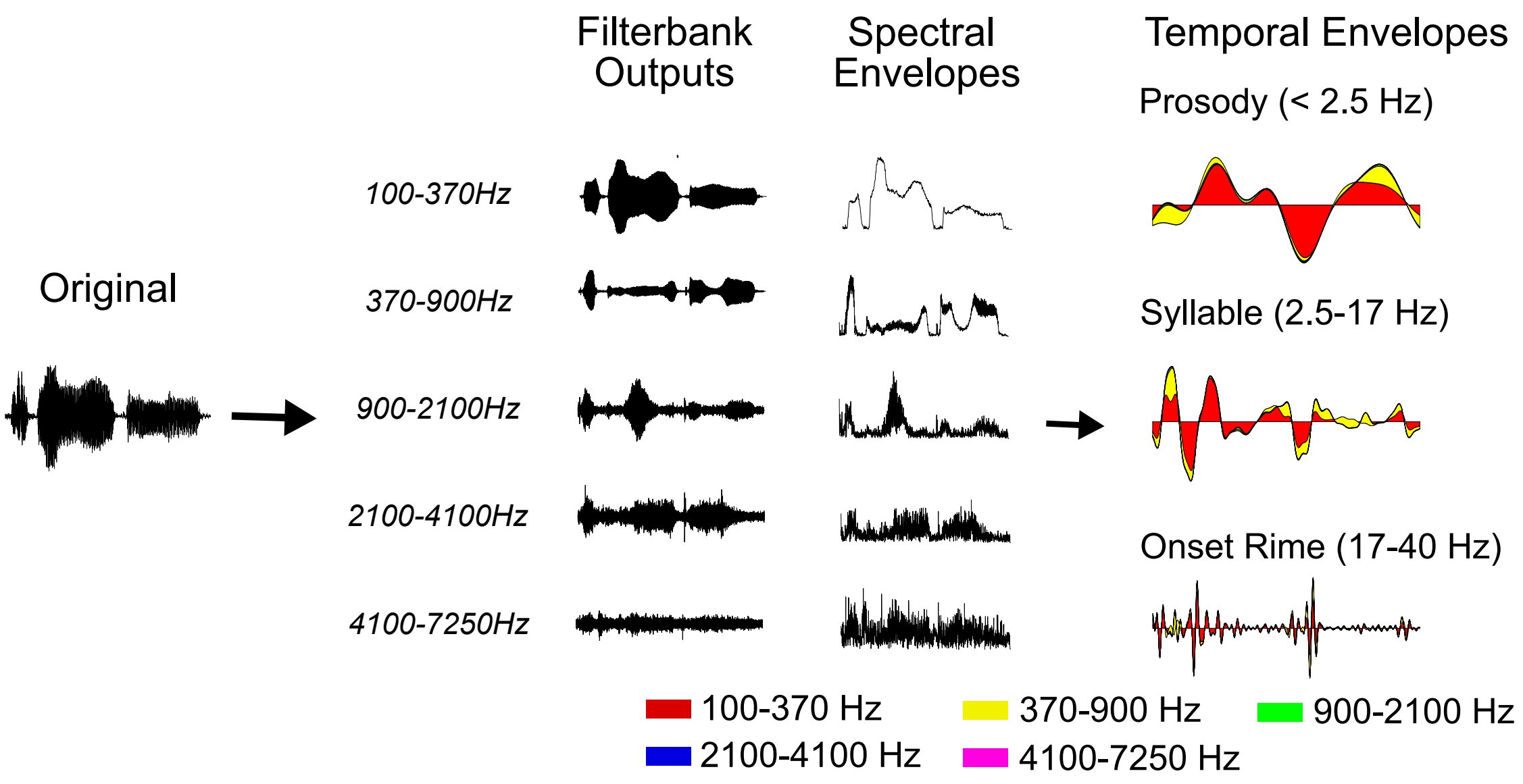

**Figure 1. Signal Processing Workflow in S-AMPH Model.** The figure illustrates the sequential processing steps of the S-AMPH model, using a sample waveform derived from this study. Step 1 involves passing the raw waveform through an $ERB_N$-spaced filter bank, covering an acoustic frequency range of 100–7,250 Hz. This process generates a hierarchical representation of the core spectral (acoustic frequency) and temporal (oscillatory rate) modulation hierarchies within the amplitude envelopes of speech, represented as Spectral Envelopes and Temporal Envelopes, respectively. The envelope for each channel is extracted using the Hilbert transform. Dimensionality reduction is achieved via PCA, which identifies patterns of covariation across spectral and temporal channels to determine the number and boundaries of modulation bands. This dimensionality reduction step facilitates the identification of core spectral and temporal modulation bands. This workflow culminates in a cascade of amplitude modulators at varying oscillatory rates, encapsulating the hierarchical structure of speech amplitude modulation.

### *2.3.2. Phase Synchronization Analyses*

We examined multi-timescale phase synchronization between temporal modulation bands by calculating the phase synchronization ratios between "adjacent" bands identified through the S-AMPH model applied to the speech samples in each language. The Phase Synchronization Index (PSI) was computed for each pair of adjacent AM bands within the S-AMPH representation, including delta vs. theta and theta vs. beta/gamma phase synchronizations. Originally developed to quantify synchronization between oscillators of differing frequencies (e.g., muscle activity; Tass et al., 1998), the n:m PSI has been adapted for analyzing neural oscillatory phase-locking (Schack & Weiss, 2005).

$$PSI = |e^{1(n\theta 1 - m\theta 2)}| \quad \ldots\ldots\ldots\ldots\ldots\ldots\ldots\ldots\ldots\ldots (1)$$

Here, n and m denote the relative frequency relationship between the lower and higher AM bands, calculated based on their cycle lengths. For example, if the cycle length of the delta rhythm is 2000 ms and the theta rhythm nested within delta has a cycle length of 1500 ms, the resulting n:m ratio is 4:3. This ratio was determined for each PSI computation. The terms θ1 and θ2 represent the instantaneous phases of the lower and higher AM bands, respectively, at each time point. The expression (nθ1 – mθ2) calculates the generalized phase difference between the two AM bands, measured as the circular distance (modulus 2π) between the instantaneous phase angles. The angled brackets indicate the mean phase difference across all time points, while the absolute value ensures the PSI remains between 0 and 1. A PSI value of 1 corresponds to perfect phase synchronization, indicative of rhythmically regular patterns perceived as a repeating sequence of strong and weak beats. Conversely, a PSI value of 0 denotes a lack of synchronization, perceived as rhythmically irregular or random patterns. This methodology enables the quantification of rhythmic coherence across temporal modulation bands, consistent with prior research (Leong et al., 2017).

# 3. Results

## 3.1. Amplitude Modulation Properties

To illustrate the overarching acoustic patterns of the speech analyzed in this study, we present scalograms displaying the exemplary frequency characteristics in Figure 2. This visualization highlights notable acoustic parallels.

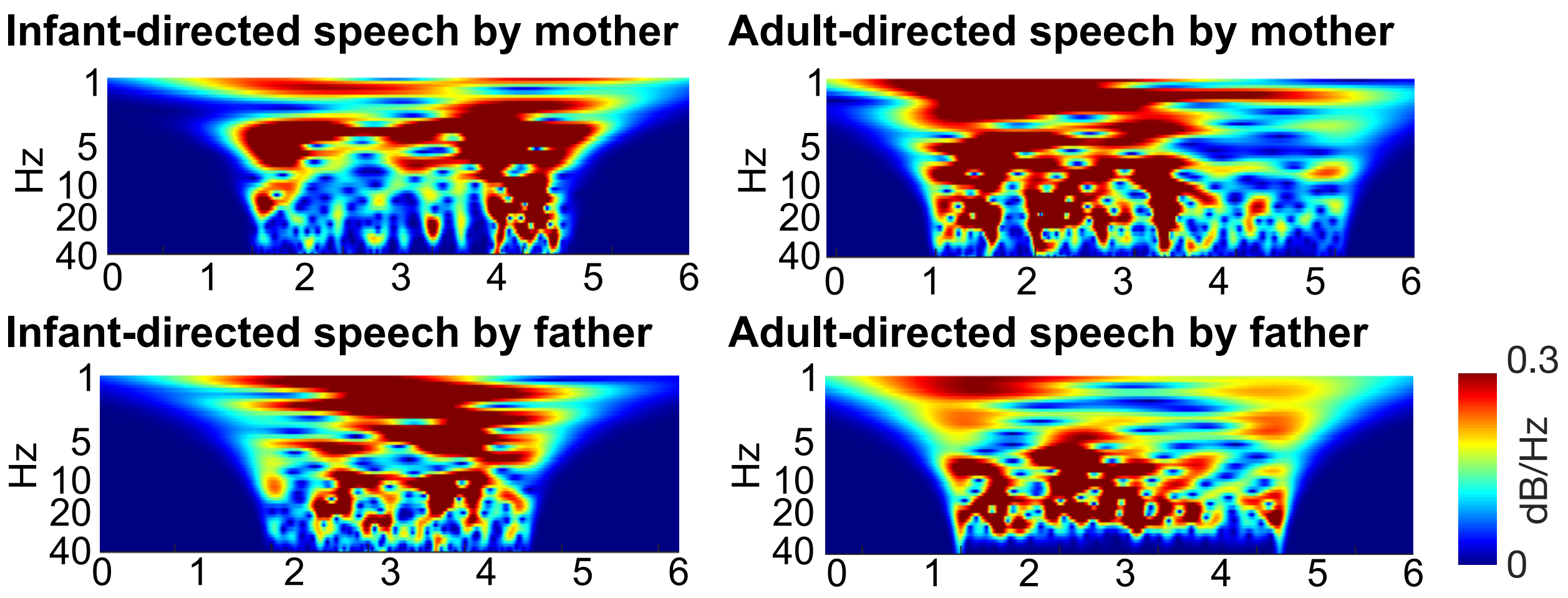

**Figure 2. Scalograms Representing the Amplitude Modulation (AM) Envelopes of Speech.** This figure presents scalograms derived from representative examples of speech samples, created using Continuous Wavelet Transform (CWT). Each scalogram is based on randomly selected 6-second excerpts of speech and depicts the AM envelopes within these samples. It is important to note that scalograms cannot be directly generated using the S-AMPH model due to the application of cochlear filter banks, which result in the loss of distinct boundary frequencies. The x-axis represents time (spanning 6 seconds), while the y-axis reflects the modulation rate (ranging from 0.1 to 40 Hz). Amplitude maxima are normalized to 0 dB, and the demodulated outputs are visualized as heat maps, capturing the modulation dynamics across time and frequency domains.

### *3.1.1. Spectral PCA*

The spectral PCA component loading patterns for each of five families were shown in the S2 Appendix. In each subplot, the lines of different thicknesses indicate different components. The loading patterns for the top 5 components are similar across the participants. It may be observed that they produced consistent loading patterns, particularly for the first 3 components (i.e., PC1, PC2, and PC3). Because of the consistency and similarity of the loading patterns across participants, the present study considered the core spectral bands using the grand average PC loading patterns. All of the contribution rates in each component and numbers of PC loading have been deposited to an external source (https://osf.io/n3upf/).

The first five principal components (PC1 to PC5) accounted for an average of 46.16%, 10.45%, 6.32%, 4.78%, and 3.86% of the total variance. The grand averages, loading patterns, and cumulative contributions for each group have been archived in an external repository (https://osf.io/n3upf/). The spectral PCA analysis uncovered distinct patterns, revealing a hierarchical organization of spectral modulation. PC1 exhibited a prominent peak around 3000 Hz, reflecting strong global coherence across spectral channels, as suggested by its alignment with average inter-channel correlation coefficients (Appendix 2). PC2 and PC 3 consistently displayed peaks near 300 Hz with "flanking" troughs around 370 Hz. These troughs demarcate the boundary separating the first and second spectral bands, providing clear evidence for the lowest spectral band within this range. Additional peaks and troughs in the loading patterns revealed the presence of four more spectral bands (Appendix S2 and Methods), resulting in distinct spectral bands identified through spectral PCA. Based on the predefined criteria outlined in the Methods, the spectral bands were delineated as follows: The five distinct bands were identified—Band 1 spanning 100–370 Hz, Band 2 encompassing 370–900 Hz, Band 3 covering 900–2100 Hz, Band 4 ranging from 2100–4100 Hz, and Band 5 extending from 4100–7250 Hz. Table 2 provides a detailed summary of the spectral bands and their boundaries.

These findings notably align with previous analyses of spectral bands in speech and music, including studies on infant-directed speech (IDS) in English, child-directed speech (CDS) in English and Spanish, adult-directed speech (ADS) in English and Portuguese, poetry readings in German and French, and a range of genres of music (Leong et al., 2017; Leong & Goswami, 2015; Araujo et al., 2018; Perez-Navarro et al., 2022; Daikoku et al., 2022; Daikoku & Goswami, 2022; Daikoku et al., 2024) which consistently identified a five-band spectral structure.

**Table 2. Summary of the spectral bands and the flanking boundaries indentified from spectral PCA of speech**

| Spectral bands | Frequency range (Hz) | PC Peaks |
|---|---|---|
| Band 1 | 100-370 | PC2-PC5 |
| Band 2 | 370-900 | PC2-PC3 |
| Band 3 | 900-2100 | PC4-PC5 |
| Band 4 | 2100-4100 | PC1 |
| Band 5 | 4100-7250 | PC2-PC5 |

### *3.1.2. Temporal PCA*

Figure 3 shows the grand average loading patterns (absolute values) for the first three principal components arising from the temporal PCA of each of the five spectral bands determined in the spectral PCA (Table a in Appendix S2). As can be observed, the loading exhibited consistent patterns among the three PCA loading patterns.

Regardless of the types of speech (ADS by mother, ADS by father, CSD by mother, CDS by father), the first to third principal components (PC1 to PC3) accounted for 71.38%, 13.00%, and 4.68% in all bands, respectively. Each group's grand average, loading patterns, and cumulative contribution was deposited to an external source (https://osf.io/n3upf/).

In all types of speech, PC1 exhibited a moderate peak at 7 Hz across all five spectral bands. Consistent with the results of the spectral PCA, PC1 in the temporal PCA likely reflects the global coherence among temporal channels, as evidenced by the absence of troughs that would signify potential boundaries. Consequently, the analysis primarily focused on PC2 and PC3. The loading patterns of PC2 revealed two prominent peaks at approximately 1.5 Hz (indicating a delta-rate AM band) and 30 Hz (indicating a beta-gamma rate AM band), alongside a pronounced flanking trough around 7 Hz. These patterns were consistent across all five spectral bands, providing strong evidence for the existence of at least two core temporal AM bands. PC3 loading patterns exhibited distinct peaks at approximately 1.5 Hz, 7 Hz, and 30 Hz. Flanking troughs for these peaks were identified at approximately 2.5 Hz and 17 Hz. Based on the pre-defined criteria (see Methods), the temporal PCA results supported the existence of three core temporal AM bands, delineated by two boundaries (Table 3). These findings align well with previous research on IDS, CDS, and music, supporting the cross-domain and cross-linguistic relevance of these temporal modulation bands (Leong et al., 2017; Leong & Goswami, 2015; Araujo et al., 2018; Perez-Navarro et al., 2022; Daikoku et al., 2022; Daikoku & Goswami, 2022). However, the theta-rate band (Band 2) is notably wider in Japanese, as in prior CDS and IDS analyses of European languages the boundary has been ~12Hz.

**Table 3. Summary of the 3 temporal bands and the 2 flanking boundaries indentified from temporal PCA**

| Temporal bands | Frequency range (Hz) | PC Peaks |
|---|---|---|
| Band 1 | 0.9-2.5 | PC2, PC3 in spectral band 1-5 |
| Band 2 | 2.5-17 | PC1, PC3 in spectral band 1-5 |
| Band 3 | 17-40 | PC2, PC3 in spectral band 1-5 |

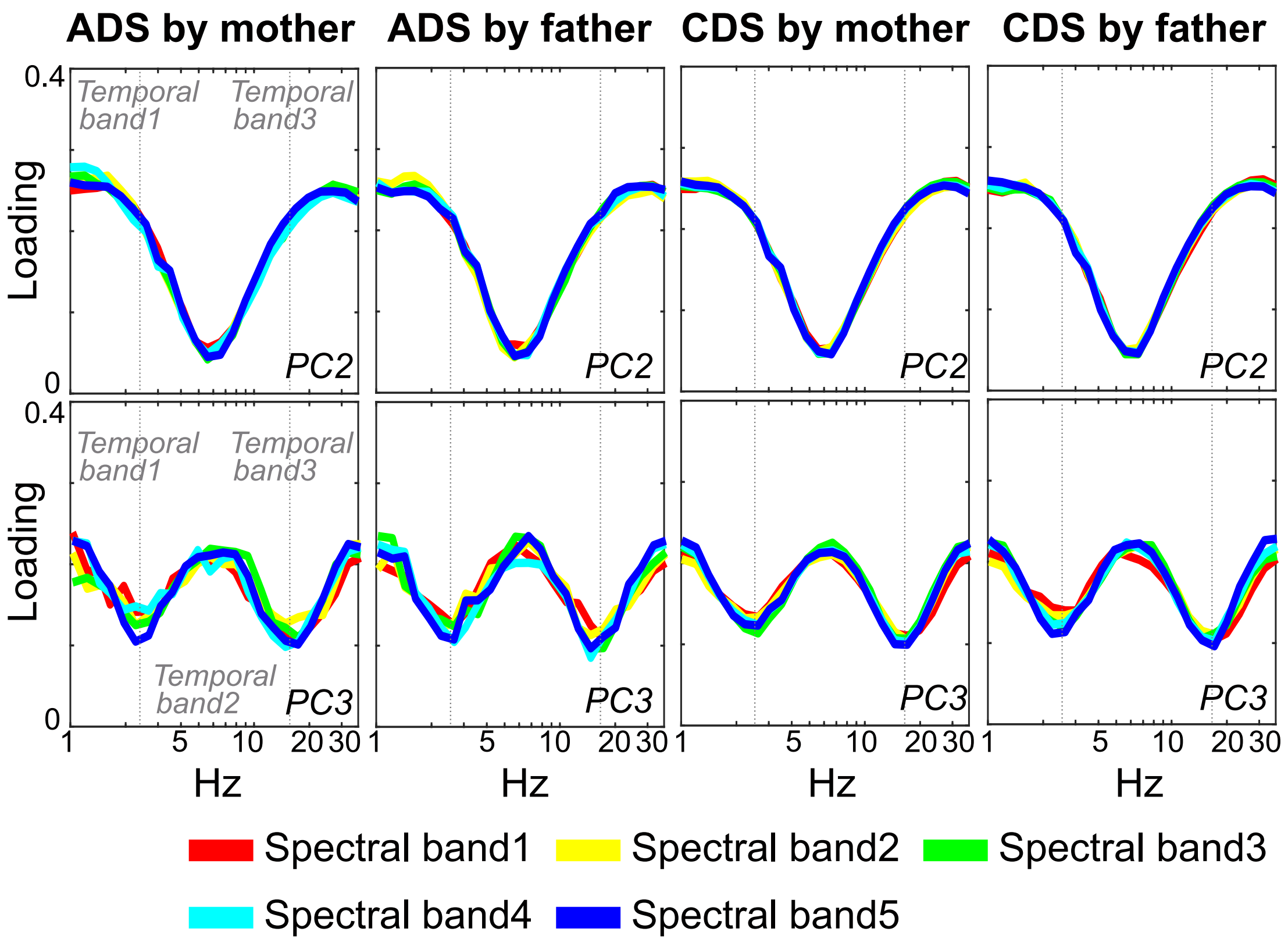

**Figure 3. Core Temporal Modulation Rates in speech.** The average absolute value of the temporal PCA component loading patterns for PC1, PC2 and PC3 generated by the S-AMPH model is depicted. The model showed an amplitude modulation (AM) hierarchy that consisted of delta-, theta- and beta/gamma-rate AM bands. Colors in the figure represent the five spectral bands.

### 3.3. Multi-Timescale Phase Synchronization

To investigate the extent of phase synchronization between modulation bands, we generated 5×3 temporal modulation envelopes corresponding to the five spectral bands and three temporal modulation bands. Notably, the mean frequency power (MFP) was significantly higher in spectral band 1 (the pitch band of the human voice) and band 2 (the pitch-to-formant transition band) compared to the other bands (S3 Appendix, Table a). This heightened power suggests that these two bands play a dominant role in shaping the rhythmic structure of speech.

We then examined the phase-synchronization indices (PSI; Leong et al., 2017) between delta and theta AM bands. Based on previous analyses of both speech and various musical genres (Leong et al., 2017; Perez-Navarro et al., 2022; Daikoku & Goswami, 2022), we anticipated that the integer ratio of 1:2 would emerge as the dominant ratio, signaling the presence of a metrical beat structure. However, in Western music (not speech) there are also dominant 1:3 and 2:3 beat structures. The PSI analyses demonstrated a high degree of consistency across participants (see Appendix 4). Further, the results revealed that the 1:2 and 1:3 integer ratios both exhibited equally high PSIs within the delta-theta AM bands. Additionally, the 2:3 integer ratios in the S-AMPH modeling showed relatively high PSIs compared to other integer ratios. While these findings align with previous research on IDS, rhythmic CDS, and music in terms of the importance of these phase synchronization patterns (Leong et al., 2017; Leong & Goswami, 2015; Perez-Navarro et al., 2022; Daikoku & Goswami, 2022), the equal dominance of 1:2 and 1:3 ratios in the current analysis of natural speech appears to be unique to Japanese. The prior studies of European languages all reported that the PSI of the 1:2 ratio was significantly stronger in IDS/CDS than in ADS. While our study revealed similar differences in PSI when comparing the 1:2 ratio in Japanese IDS/CDS with Japanese ADS, with significantly stronger rhythmic coupling in IDS/CDS ($p < 0.05$, Figures 4b and 4c), we found similar differences between ADS and IDS/CDS for the 1:3 ratio ($p < 0.05$) and for the 2:3 ratio ($p < 0.05$) (for details, https://osf.io/n3upf/). This is suggestive of additional AM-phase relations, which may possibly yield moraic rhythms. Further, within IDS/CDS, a decline in the PSI was observed as the child's age increased (see Figure 4d) (all: $p < 0.05$). This suggests developmental changes in how prosodic features align with syllabic structures over time, and was not previously tested for European languages as all prior data sets were cross-sectional.

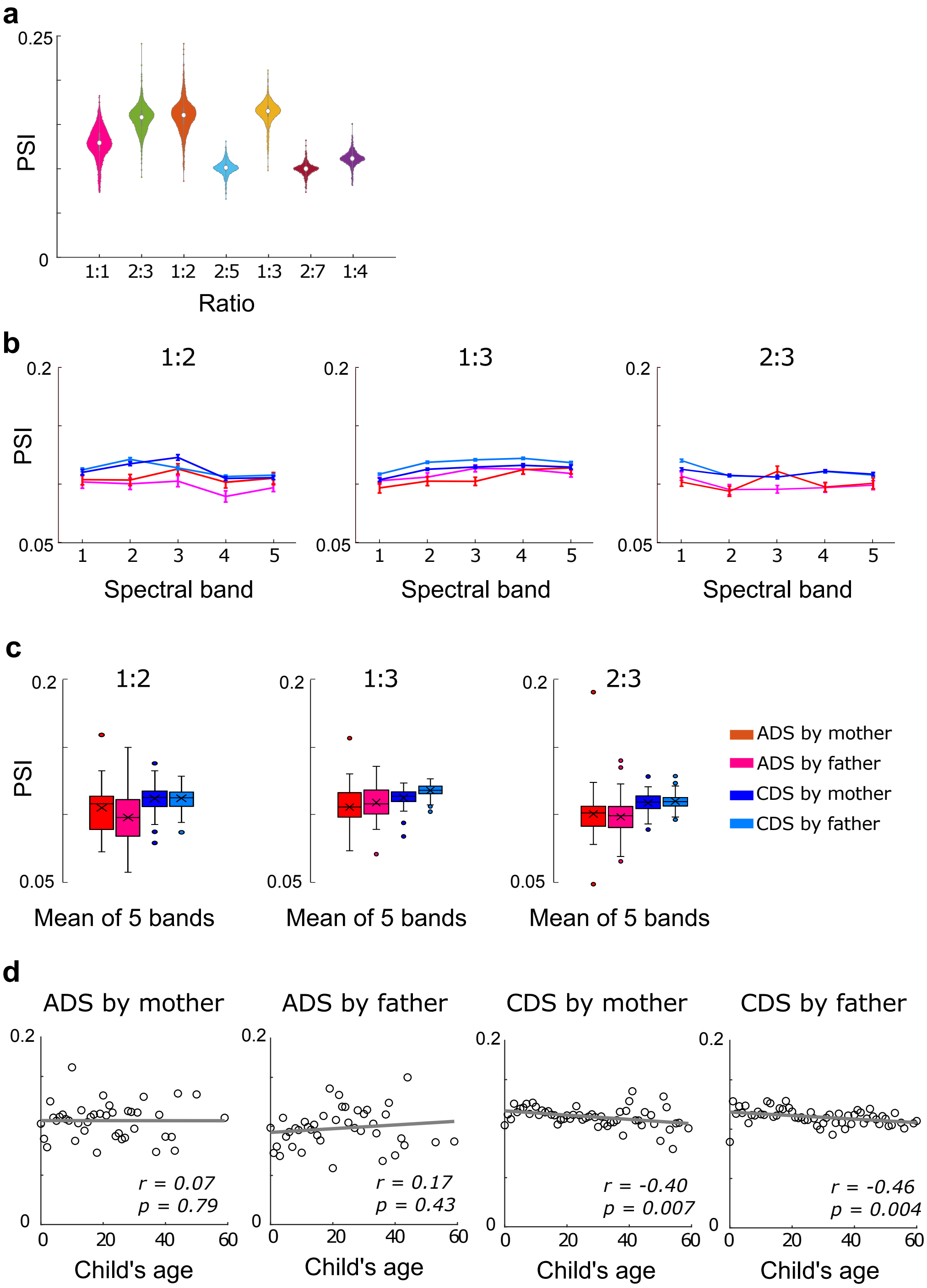

**Figure 4. Phase synchronization index (PSI) for each type of speech.** Each integer ratio (a) in PSI averaged across the 5 spectral bands. The simpler integer ratios 1:2 and 1:3 ratios are as prevalent as each other in this corpus, and the 2:3 ratio is also prominent. The 2:3 ratio may however reflect a mathematical averaging effect regarding 1:2 and 1:3. The PSI in the 1:2 ratio was significantly stronger in CDS compared to ADS (b and c).

Additionally, within CDS, a decline in the PSI was observed as the child's age increased (d).

# 4. Discussion

Here we applied a language-blind computational model of linguistic rhythm based on features of the amplitude envelope to compute the physical acoustic stimulus characteristics that characterize Japanese IDS, CDS and ADS. The Spectral-Amplitude Modulation Phase Hierarchy (S-AMPH) model identified systematic temporal modulation patterns and phase synchronization indices (PSIs) across these three speech types, which largely matched prior findings in European syllable-timed (French, Spanish, Portuguese) and stress-timed (English, German) languages (Leong & Goswami, 2015; Araujo et al., 2018; Daikoku & Goswami, 2022; Perez-Navarro et al., 2022; Daikoku et al., 2024). As in these European languages, the modelling demonstrated 5 spectral bandings and 3 temporal bandings, with very similar ranges in Hz irrespective of whether the Japanese mother or the Japanese father was the speaker. These systematic temporal modulation patterns would enable cortical speech tracking by oscillatory networks operating at the rates of delta, theta and beta/low gamma in whichever language environment an infant was born into. As also found in European languages, the delta-theta PSIs were significantly higher in speech directed to infants and children than in speech directed to adults. This suggests a language-teaching function regarding these AM bandings and their PSIs. This is supportive of the production of universal acoustic physical structures by speakers of IDS and CDS, that may serve as a foundational perceptual scaffold for infant language development.

However, the modelling also demonstrated some differences between IDS and CDS in Japanese and previously-modelled European languages. One important cross-language difference was that the theta-rate AM banding in Japanese IDS/CDS was elongated, spanning 2.5 – 17Hz. This contrasts with the theta-rate AM band width in English IDS and CDS (stress-timed rhythm, 2.5-12 Hz), and Spanish CDS (syllable-timed rhythm, 2.5-12 Hz). It can be speculated that this elongation enables the use of moraic units to produce rhythmic regularity in Japanese. Moraic units are thought to occur at ~10 Hz (see Peter et al., 2022). A wider theta banding would enable the Japanese-learning infant brain to accommodate moraic parsing. A second important cross-language difference was that 1:3 ratios were as frequent as 1:2 ratios when the PSIs were computed for the two slowest AM bandings, the delta-rate and theta-rate AM bandings. Rather than the perception of rhythmic "beat" being driven only by a metronome-like 1:2 ratio, it was equally likely to be driven by a 1:3 ratio. A PSI of 1:2 indicates a clear metrical beat with two syllables per foot, as in "JACK and JILL went UP the HILL" (capitalized syllables indicate greater syllabic stress). A PSI of 1:3 means that there are also triplet beats (3 syllables per foot, an English example would be "PUSS-y-cat PUSS-y-cat, WHERE have you BEEN?), which can presumably accommodate moraic syllabic parsing such as in the example PA-N-DA. One possible conclusion is that Japanese-learning infants are extracting both kinds of rhythmic structure from IDS simultaneously. When modelling Western music (jazz, classical, rock; Daikoku & Goswami, 2022), 1:3 ratios were also prominent, supporting the apparent universality of AM-ratio structures in governing rhythmic patterning. The stronger 2:3 ratio observable in Figure 4a is likely a mathematical effect, reflecting an averaging of the 1:2 and 1:3 ratios.

The finding that Japanese-learning infants can parse both syllabic and moraic units from the speech stream early in life is supported by prior work showing that Japanese infants' use of moraic units is weaker than English- and French-learning infants' use of syllable and stress information (Sato et al., 2010, 2012). Sato and her colleagues have demonstrated that Japanese-learning infants do not discriminate monomoraic from bimoraic syllables until 10 months of age (Sato et al., 2010, 2012). Accordingly, the mora in Japanese may be a less prominent rhythmic unit than the foot or the syllable. Further, in the recent adult cortical speech tracking study reported by Peter et al. (2022), synthetic speech was created that could be parsed at the level of the foot (2.5 Hz), syllable (5 Hz) or mora (10 Hz). When EEG was recorded while English, French and Japanese listeners were tested, the Japanese adults did not show preferential tracking at 10Hz as had been predicted on a language-specific hypothesis (Peter et al., 2022). Rather, all three groups showed the highest response amplitudes at 5 Hz, the syllable rate.

Finally, and as also found in previous studies of IDS, CDS and ADS in European languages, notable differences in the strength of PSIs were observed between Japanese IDS/CDS and Japanese ADS. Japanese IDS/CDS exhibited significantly stronger phase synchronization than Japanese ADS. This elevated delta-theta PSI contributes to the exaggerated prosodic features of infant- and child-directed speech, providing the learning brain with more salient rhythmic acoustic cues (Leong et al., 2017; Amano et al., 2006; Perez-Navarro et al., 2022). Within Japanese IDS/CDS, we also observed a decline in PSI as children aged. This is the first longitudinal dataset modelled using the S-AMPH, and this developmental trend is not unexpected. It suggests that as children become more capable of parsing speech and producing language, the prosodic emphasis in caregiver speech diminishes, moving gradually towards the acoustic physical structures that characterize ADS. Importantly, however, the data analysed here were recordings of natural conversational speech. In deliberately rhythmic speech, such as poetry and proverbs, ADS continues to exhibit significantly stronger delta-theta PSIs than conversational speech (Araujo et al., 2018).

The longitudinal corpus used in this study has also been utilized in earlier investigations of Japanese infant-directed speech, thereby demonstrating that all the different language-teaching characteristics of IDS are present at the same time (e.g., heightened pitch, exaggerated pitch range and hyperarticulation of vowels, Amano et al., 2006, Ishizuka et al., 2007). Ishizuka et al. (2007) demonstrated developmental changes in the spectral peaks of vowels, while Amano et al. (2006) conducted a detailed analysis of the fundamental frequency (F0) of IDS, reporting that F0 decreases nearly linearly with infant age. Amano et al. also found that in IDS, the F0 is significantly higher than in ADS during the period from birth to around 18 months, the age at which infants typically begin producing two-word utterances. Over this period, F0 in Japanese IDS decreases as the child ages. Our findings provide complementary evidence of an accompanying decline in the phase synchronization between the delta-rate and theta-rate AM bandings. Via our novel focus on AM hierarchies, we demonstrate how the different language-teaching characteristics of IDS change over a similar time-frame.

One limitation of this study is the small sample size of six speakers, dictated by the logistical challenges of obtaining long-term naturalistic recordings over five years. Although longitudinal studies provide unparalleled insights into individual developmental trajectories, the small number of speakers may limit the generalizability of the findings. Nevertheless, the first study to apply the S-AMPH model to English CDS also used 6 speakers (Leong & Goswami, 2015), and identified temporal and spectral bandings which were replicated in a later study with 24 speakers (Leong et al., 2017). Future research could aim to replicate the current study using larger, cross-sectional datasets. Additionally, while the AM hierarchy and PSI analyses revealed robust patterns, further studies could examine cross-linguistic comparisons involving other mora-timed languages, such as Gilbertese or Slovak, to better understand whether the rhythmic characteristics observed here are unique to Japanese or reflect broader linguistic typologies.

In conclusion, this study contributes to the growing body of research exploring the rhythmic properties of speech, and the special rhythmic characteristics of infant-directed and child-directed speech. By uncovering broadly similar AM hierarchies and PSI patterns in Japanese speech as found in European stress-timed and syllable-timed languages, we provide evidence for the universality of rhythmic structures in human communication while highlighting the temporal dynamics unique to Japanese IDS/CDS. These findings deepen our understanding of how universal physical acoustic structures enable cortical tracking across languages while simultaneously accommodating linguistic differences in rhythm type, offering unique insights into the fundamental mechanisms of language acquisition.

# 5. References

Amano, S., Kondo, T., and Kato, K. (2008). NTT Infant Speech Database (INFANT). Speech Resources Consortium, National Institute of Informatics. (dataset). https://doi.org/10.32130/src.INFANT

Araújo, J., Flanagan, S., Castro-Caldas, A., & Goswami, U. (2018). The temporal modulation structure of illiterate versus literate adult speech. *PLOS ONE*, *13*(10), e0205224. https://doi.org/10.1371/journal.pone.0205224

Daikoku T, Goswami U. Hierarchical amplitude modulation structures and rhythm patterns: Comparing Western musical genres, song, and nature sounds to Babytalk. PLoS One. 2022;17: e0275631. doi: 10.1371/journal.pone.0275631.

Daikoku, T., Lee, C., & Goswami, U. (2024). Amplitude modulation structure in French and German poetry: universal acoustic physical structures underpin different poetic rhythm structures. Royal Society Open Science, 11(9), 232005.

Dau, T., Kollmeier, B., & Kohlrausch, A. (1997). Modeling auditory processing of amplitude modulation I. Detection and masking with narrow-band carriers. *Journal of the Acoustical Society of America, 102*(5 Pt 1), 2892–2905. https://doi.org/10.1121/1.420344

Dauer, R. M. (1983). Stress-timing and syllable-timing reanalyzed. Journal of Phonetics, 11(1), 51–62.

Fitch, W. T. (2012). The biology and evolution of rhythm: Unraveling a paradox. Language and music as cognitive systems, 73-95.

Giraud, A. L., & Poeppel, D. (2012). Cortical oscillations and speech processing: emerging computational principles and operations. Nature neuroscience, 15(4), 511-517.

Leong, V. (2012). *Prosodic rhythm in the speech amplitude envelope: Amplitude modulation phase hierarchies (AMPHs) and AMPH models* [PhD Thesis].

Leong, V., Kalashnikova, M., Burnham, D., & Goswami, U. (2017). The temporal modulation structure of infant-directed speech. *Open Mind*, *1*(2), 78–90. https://doi.org/10.1162/OPMI_a_00008

Leong, V., & Goswami, U. (2015). Acoustic-emergent phonology in the amplitude envelope of child-directed speech. *PLOS ONE, 10*(12), e0144411. https://doi.org/10.1371/journal.pone.0144411

Lerdahl, F. (2001). The sounds of poetry viewed as music. Annals of the New York Academy of Sciences, 930(1), 337-354.

Mehler, J., Jusczyk, P., Lambertz, G., Halsted, N., Bertoncini, J., & Amiel-Tison, C. (1988). A precursor of language acquisition in young infants. *Cognition*, *29*(2), 143–178. https://doi.org/10.1016/0010-0277(88)90035-2

Moore, B. C. J. (2012). *An introduction to the psychology of hearing*. Brill.

Nazzi, T., Bertoncini, J., & Mehler, J. (1998). Language discrimination by newborns: Toward an understanding of the role of rhythm. *Journal of Experimental Psychology: Human Perception and Performance*, *24*, 756-766. https://doi.org/10.1037/0096-1523.24.3.756

Patel, A. D. (2003). Rhythm in language and music: parallels and differences. Annals of the New York Academy of Sciences, 999(1), 140-143.

Grabe, E., & Low, E. L. (2002). Durational variability in speech and the rhythm class hypothesis. In C. Gussenhoven & N. Warner (Eds.), *Laboratory Phonology 7* (pp. 515-546). Mouton de Gruyter.

Greenberg, S. (2006). A multi-tier framework for understanding spoken language. In S. Greenberg & W. Ainsworth (Eds.), *Listening to speech: An auditory perspective*. Lawrence Erlbaum Associates.

Gross, J., Hoogenboom, N., Thut, G., Schyns, P., Panzeri, S., Belin, P., & Garrod, S. (2013). Speech rhythms and multiplexed oscillatory sensory coding in the human brain. *PLOS Biology*, *11*(12), e1001752. https://doi.org/10.1371/journal.pbio.1001752

Klein, W., Plomp, R., & Pols, L. C. W. (1970). Vowel spectra, vowel spaces, and vowel identification. *Journal of the Acoustical Society of America*, *48*(4), 999–1009. https://doi.org/10.1121/1.1912239

Kotz, S. A., Ravignani, A., & Fitch, W. T. (2018). The evolution of rhythm processing. Trends in cognitive sciences, 22(10), 896-910.

Perez-Navarro, J., Lallier, M., Clark, C., Flanagan, S., & Goswami, U. (2022). Local temporal regularities in child-directed speech in Spanish. Journal of Speech, Language and Hearing Research, 65(10), 3776-3788.

Poeppel, D., & Assaneo, M. F. (2020). Speech rhythms and their neural foundations. Nature reviews neuroscience, 21(6), 322-334.

Pols, L. C. W., Tromp, H. R. C., & Plomp, R. (1973). Frequency analysis of Dutch vowels from 50 male speakers. *Journal of the Acoustical Society of America*, *53*(4), 1093–1101. https://doi.org/10.1121/1.1913429

Schack, B., & Weiss, S. (2005). Quantification of phase synchronization phenomena and their importance for verbal memory processes. *Biological Cybernetics*, *92*(4), 275–287. https://doi.org/10.1007/s00422-005-0555-1

Stone, M. A., & Moore, B. C. J. (2003). Tolerable hearing aid delays. III. Effects on speech production and perception of across-frequency variation in delay. *Ear and Hearing*, *24*(2), 175–183. https://doi.org/10.1097/01.AUD.0000058106.68049.9C

# Supporting information

S1 Appendix. Signal Processing Steps in S-AMPH Model.
S2 Appendix. Individual Variation of PCA Loadings in S-AMPH model.
S3 Appendix. Mean Power of 5 spectral Envelopes, the FFT results, and Individual Variation of PSI in Each Integer Ratio.